\documentclass[aps,pra,twocolumn,superscriptaddress]{revtex4-1}
\usepackage{mathrsfs}
\usepackage{epsfig}
\usepackage{graphicx}
\usepackage{amsfonts}
\usepackage[figuresright]{rotating}
\usepackage{amssymb}
\usepackage{amsmath}
\usepackage{dcolumn}
\usepackage{bm}
\usepackage{color}
\usepackage[colorlinks=true,linkcolor=blue,citecolor=blue,anchorcolor=green,urlcolor=blue]{hyperref}
\usepackage{placeins}

\def\be{\begin{equation}}
\def\ee{\end{equation}}
\def\bea{\begin{eqnarray}}
\def\eea{\end{eqnarray}}

\def\ba{\begin{aligned}}
\def\ea{\end{aligned}}

\newcommand{\WQC} {Wilczek Quantum Center and Key Laboratory of Artificial Structures and Quantum Control, School of Physics and Astronomy, Shanghai Jiao Tong University, Shanghai 200240, China}

\begin{document}
\title{Multiple intermediate phases in the interpolating Aubry-Andr\'{e}-Fibonacci model }

\author{Chenyue Guo}
\email{guochenyue@sjtu.edu.cn}
\affiliation{\WQC}


\begin{abstract}
We investigate a generalized interpolating Aubry-Andr\'{e}-Fibonacci (IAAF) model with p-wave superconducting pairing, focusing on its localization and topological properties. Within the Aubry-Andr\'{e} limit, we demonstrate that the system experiences transitions from a pure phase, either extended or critical, to a variety of intermediate phases and ultimately enters a localized phase with increasing potential strength. These intermediate phases include those with coexisting extended and localized states, extended and critical states, localized and critical states and a mix of extended, critical and localized states.
Each intermediate phase exhibits at least one type of mobility edge separating different states. 
As the system approaches the Fibonacci limit, both the extended and localized phases diminish, and the system tends towards a critical phase.
Furthermore, the model undergoes a transition from topologically nontrivial to trivial phase as potential strength increases.
\end{abstract}
\maketitle

\section{introduction}
\label{sec1}
The study of quantum localization plays an important role in condensed matter physics, particularly since the remarkable discovery of Anderson localization in 1958~\cite{Anderson1958}. 
It indicates the absence of the delocalization-localization phase transition in low-dimensional disordered systems
~\cite{Anderson1979, Andersonreview1.1985, Andersonreview2.2008}.
Later, quasiperiodic (QP) potentials have garnered considerable attention for enabling localization transitions in one-dimensional (1D) systems.
These potentials have been successfully implemented in various experimental platforms, such as in photonic crystals 
~\cite{experimental.LY.2009, experimental.KYE.2012, experimental.Wang2020}, 
ultracold atoms~\cite{experimental.Roati2008, experimental.Modugno_2010} and so on~\cite{experimental.62.977, experimental.2003, experimental.93.053901}.
The Aubry-Andr\'{e} (AA) model~\cite{Aubry1980} stands out by demonstrating a phase transition from an extended to a localized phase when the quasiperiodic disorder strength exceeds a critical threshold. 
%
Similarly, the Fibonacci model, known for its eigenstates that remain critical at all potential strengths, has garnered considerable theoretical
~\cite{Fib.K.1983, Fib.O.1983, Fib.MR.1985, Fib.AJA.1989, Fib.RS.1997, Fib.ME.1999, Fib.FD.2001, Fib.MN.2017, Fib.JA.2021, Fib.RA.2023}
and experimental
~\cite{Fib.RS.1997, Fib.exp.DN.2003, Fib.exp.TD.2014, Fib.exp.B.2017, Fib.exp.R.2023} interest.
%
Both models belong to the same topological class and are regard as two distinctive limits within the interpolating Aubry-Andr\'{e}-Fibonacci (IAAF) model~\cite{IAAF.KYE.2012,
IAAF.VMZ.2013, IAAF.VMZ.2015}. The IAAF model provides a unique playground for investigating the localization properties~\cite{IAAF.Goblot2020, IAAF.Zhai.2021, IAAF.MBL.2021, IAAF.Dai.2023}.
For instance, Ref~\cite{IAAF.Goblot2020, IAAF.Dai.2023} present
various cascade behaviors of eigenstates during the continuous transformation of the AA model into the Fibonacci model.

The concept of mobility edge is crucial in separating extended from localized states, leading to many novel insights in fundamental physics~\cite{Andersonreview2.2008, MobilityEdge.WRS.2014, MobilityEdge.YKA.2017, MobilityEdge.CC.2020}. 
The quantum phase where extended and localized states coexist within the energy spectrum is termed the intermediate phase. 
Numerous theoretical studies have confirmed the existence of this intermediate phase and the mobility edge in one-dimensional systems with broken self-duality symmetry~\cite{MobilityEdge1.1988, MobilityEdge4.2010, MobilityEdge3.2009, MobilityEdge9.2020, MobilityEdge10.2021, MobilityEdge11.2023, MobilityEdge2.1990, MobilityEdge5.2015, 
MobilityEdge6.2017, MobilityEdge7.2018, MobilityEdge72.2019, MobilityEdge8.2020, MobilityEdge12.2023}.  
%
In contrast to phases where all eigenstates are exclusively extended or localized, there exists a distinct third phase, known as the critical phase, where all eigenstates are extended but nonergodic, as observed in generalized quasiperiodic models~\cite{critical_1990, critical_2004, pair.Gao.2016, critical_2015}.
%
Further studies~\cite{intermediate.WYC.2022, intermediate.Shilpi.2023, intermediate.Lin.2023, critical.Li.2023}
have identified an anomalous mobility edge separating the extended and localized states from the critical ones. 
These findings indicate the existence of additional intermediate phases  there is a coexistence of critical and other states.

Topological phases of matter have emerged as a fascinating area of research in condensed matter physics, offering novel insights into the behavior of quantum systems. 
These phases are characterized by their robust properties, which are protected by topological invariants and are insensitive to local perturbations\cite{MZM_2017_Wen, MZM_2010_Hasan, MZM_2011_Qi, MZM_2005_Kane, MZM_2007_Fu, MZM_2007_Moore}. 
Among the most intriguing features of topological phases are the presence of exotic quasiparticle excitations known as Majorana zero modes (MZMs)\cite{MZM_2001_Kitaev, MZM_2000_R, MZM_2001_Ivanov}.
MZMs arise in certain topological superconductors and are predicted to exist at the ends of one-dimensional systems or in vortices of two-dimensional systems\cite{MZM_2008_Nayak, MZM_2010_Lutcgyn, MZM_2010_Oreg}.
The unique properties of Majorana zero modes, such as their ability to encode and manipulate quantum information in a fault-tolerant manner, have attracted significant attention from both theoretical and experimental perspectives\cite{MZM_2014_Stevan, MZM_2016_MT, MZM_2016_ASM}.

In this paper, we explore a generalized quasiperiodic model, namely the IAAF model with p-wave superconducting (SC) pairing terms. 
We find that the potential effectively transforms into a cosine QP modulation up to a constant on-site chemical potential shift in the AA limit (see Fig.~\ref{fig.1}).
The system undergoes transitions from a pure phase, either extended or critical, to a localized phase with a strong enough potential strength. Many types of intermediate phases emerged during this process, including those with coexisting extended and localized states, extended and critical states, localized and critical states and a coexistence of extended, critical and localized states.
Specially, each intermediate phase exhibits at least one type of mobility edge separating different states.
As the system approaches the Fibonacci limit where the potential corresponds to a step potential switching between $\pm1$ values according to the Fibonacci substitution rule (see Fig.~\ref{fig.1}), the domains for extended and localized phases diminish, leading the system towards a critical phase.
Additionally, we observe that the model experiences a transition from topologically nontrivial to trivial phase via increasing the strength of potential.
MZMs are present in the topological nontrivial phase.

The structure of the paper is as follows: 
In Sec. \ref{sec2}, we briefly introduce the Bogoliubov-de Gennes (BdG) theory and outline several physical quantities to characterize the extended, localized and critical states, as well as the corresponding phases.
In Sec. \ref{sec3} and Sec. \ref{sec4}, we study the localization properties ranging from AA limit to Fibonacci limit. In Sec. \ref{sec5}, we analyse the topological properties of our model. Sec. \ref{sec6} provides the conclusion and outlook.

\section{Model and method}
\label{sec2}
Here, we start from the 1D p-wave superconducting paired IAAF
model with Hamiltonian defined as
\be 
\label{1}
\hat{H}=\sum_{i}{[-J\hat{c}_i^\dagger c_{i+1}+
\Delta \hat{c}_{i}\hat{c}_{i+1}+h.c.+
\lambda V_{i}(\beta)\hat{n}_{i}]},
\ee
where $i$ denote the lattice site index. 
$\hat{c}_{i}$($\hat{c}_{i}^\dagger$) is annihilation (creation) operator of the spinless fermion on $i$ and 
$\hat{n}_{i}=\hat{c}_{i}^\dagger \hat{c}_{i}$.
$J$ is the nearest-neighboring (NN) single-particle hopping amplitude and let $J=1$ in this paper.
$\Delta$ is the pair-driving rate, which we take as
real and positive. 
$\lambda$ is the strength of the quasiperiodically modulated 
on-site chemical potential.
The potential $V_i$ reads
\be 
\label{2}
V_i(\beta)=-\frac{\rm{tanh}[\beta(cos(2\pi \alpha \emph{i}+\theta)-cos(\pi \alpha))]}
{\rm{tanh}\beta}.
\ee

\begin{figure}
\centering
 \includegraphics[width=8.5cm]{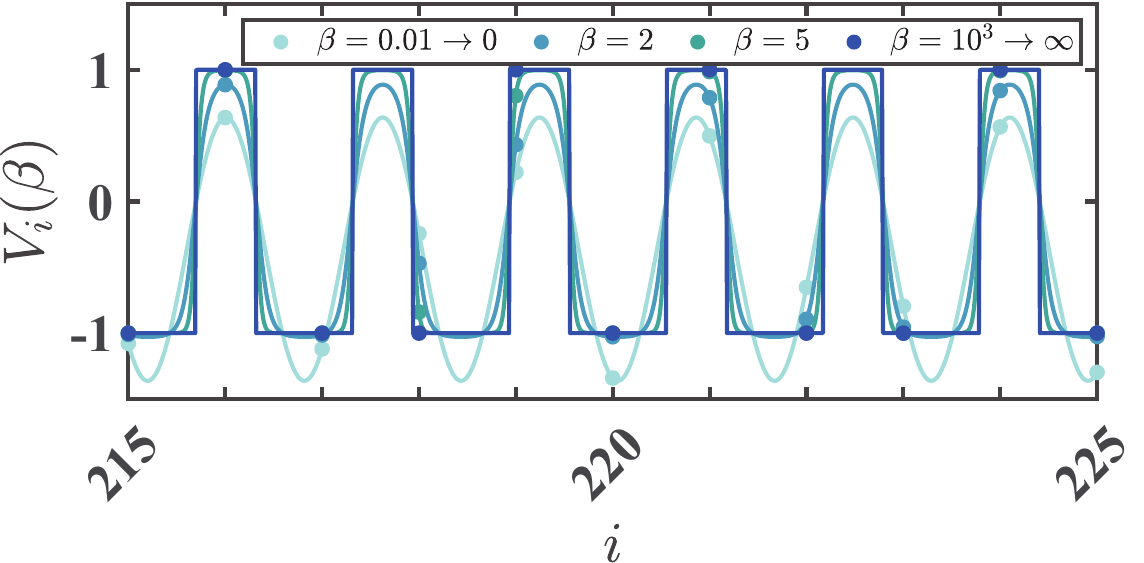}
 \caption{(Color online). Schematic of the quasiperiodically modulated 
on-site potential (eq. (\ref{2})) for several values of $\beta=0.01$, $0.2$, $5$, $1000$ (light to deep blue curves).}
 \label{fig.1}
\end{figure}

Without loss of generality, we set the phase term of the potential to be zero
($\theta=0$).
The golden mean ratio $\alpha$ can be derived from the limit of the ratio of consecutive Fibonacci numbers $F_i$~\cite{Fn.KM.1983}:
$\alpha=\lim_{n\rightarrow \infty}\frac{F_{n-1}}{F_n}=\frac{\sqrt{5}-1}{2}$
with $F_0=F_1=1$.
The parameter $\beta$ serves as a control mechanism allowing interpolation between two known limiting cases of $\beta\rightarrow0$ 
and $\beta\rightarrow\infty$.
For the former, the potential $V_i(\beta)$ simplifies to 
${\rm{cos}}(2\pi \alpha i+\theta)-{\rm{cos}}(\pi \alpha)$.
Then the model becomes a $1$D p-wave superconductor in the
incommensurate lattices~\cite{pair.Cai.2013}
up to a constant on-site chemical potential shift.
For the latter, $V_{i}(\beta)$ corresponds to a step potential switching between $\pm1$ values following the Fibonacci substitution rule.   
Fig.~\ref{fig.1} illustrates the on-site potential $V_{i}(\beta)$ to have a more intuitive understanding.

Considering the Hamiltonian. (\ref{1}) owns particle-hole symmetry, we can employ the Bogoliubov-de Gennes (BdG) transformation~\cite{BdG.KM.1980} to diagonalize it, as follows:
\be 
\label{3}
\hat{\gamma}_{\mu}^\dagger=\sum_{i=1}^{L}({v_{i\mu}\hat{c}_{i}+
u_{i\mu}\hat{c}_{i}^\dagger}),
\ee
where $L$ is the number of lattice sites and $\mu=1,...,L$.
In this paper, we set $L=F_{n-1}/F_n$ to ensure a periodic boundary condition.
Then the eq. (\ref{1}) in terms of the $\hat{\gamma}_\mu$ and 
$\hat{\gamma}_\mu^\dagger$ operators reads: 
\be 
\label{4}
\hat{H}=\sum_{\mu=1}^{L}2\epsilon_\mu
(\hat{\gamma}_\mu^\dagger\hat{\gamma}_\mu-\frac{1}{2}).
\ee
Assuming the energy spectrum $\epsilon_\mu$ is non-negative. The eigenstates in terms of spinless fermion language is defined as 
$|\Phi_\mu\rangle=(u_\mu,v_\mu)^T=(u_{\mu 1},...,u_{\mu L},v_{\mu 1},...v_{\mu L})^{T}$ 
and the positive eigenvalues $\epsilon_\mu$ are obtained by solving Bogoliubov-de Gennes equation:
\begin{align}
\label{5}
\mathcal{\hat{H}}
\left(\begin{array}{cc}
u_{\mu}\\
v_{\mu}
\end{array}\right)
=\left(\begin{array}{cc}
A & B\\
-B^* & -A^*
\end{array}\right)
\left(\begin{array}{cc}
u_{\mu}\\
v_{\mu}
\end{array}\right)
=\epsilon_\mu
\left(\begin{array}{cc}
u_{\mu}\\
v_{\mu}
\end{array}\right).\nonumber\\
\end{align}

Given that all couplings are real in our model, the associated
$2L\times2L$ matrices $\mathcal{\hat{H}}$ is real and symmetric. 
Hence the matrix $A$ is real and symmetric $(A=A^*=A^T)$, 
while $B$ is real and anti-symmetric $(B=B^*=-B^T)$. The eq. (\ref{5}) can be further read as:
\begin{align}
\label{52}
&(A+B)\phi_\mu =\epsilon_\mu \psi_\mu, 
&(A-B)\psi_\mu =\epsilon_\mu \phi_\mu,
\end{align}

where $\phi_\mu=u_\mu+v_\mu$ and $\psi_\mu=u_\mu-v_\mu$.
The elements of $\mathcal{\hat{H}}$ are defined as $2A_{ij}=\lambda V_{ij}\delta_{ij}-J(\delta_{j,i+1}+\delta_{j,i-1})$,
$2B_{i,j}=-\Delta(\delta_{j,i+1}-\delta_{j,i-1})$.
The eigenvector components are defines as
$u_\mu^T=(u_{\mu 1},...,u_{\mu L})^{T}$ and
$v_\mu^T=(v_{\mu 1},...,v_{\mu L})^{T}$.
The eigenvalues satisfy $\hat{\gamma}_\mu(\epsilon_\mu)=
\hat{\gamma}_\mu^\dagger(-\epsilon_\mu)$ where only the zero-energy states 
$(\epsilon_\mu=0)$ are self-conjugate due to the particle-hole symmetry.
Our calculations will focus solely on the quasiparticle spectra of the BdG Hamiltonian for simplify.

\begin{figure}
\centering
 \includegraphics[width=8.5cm]{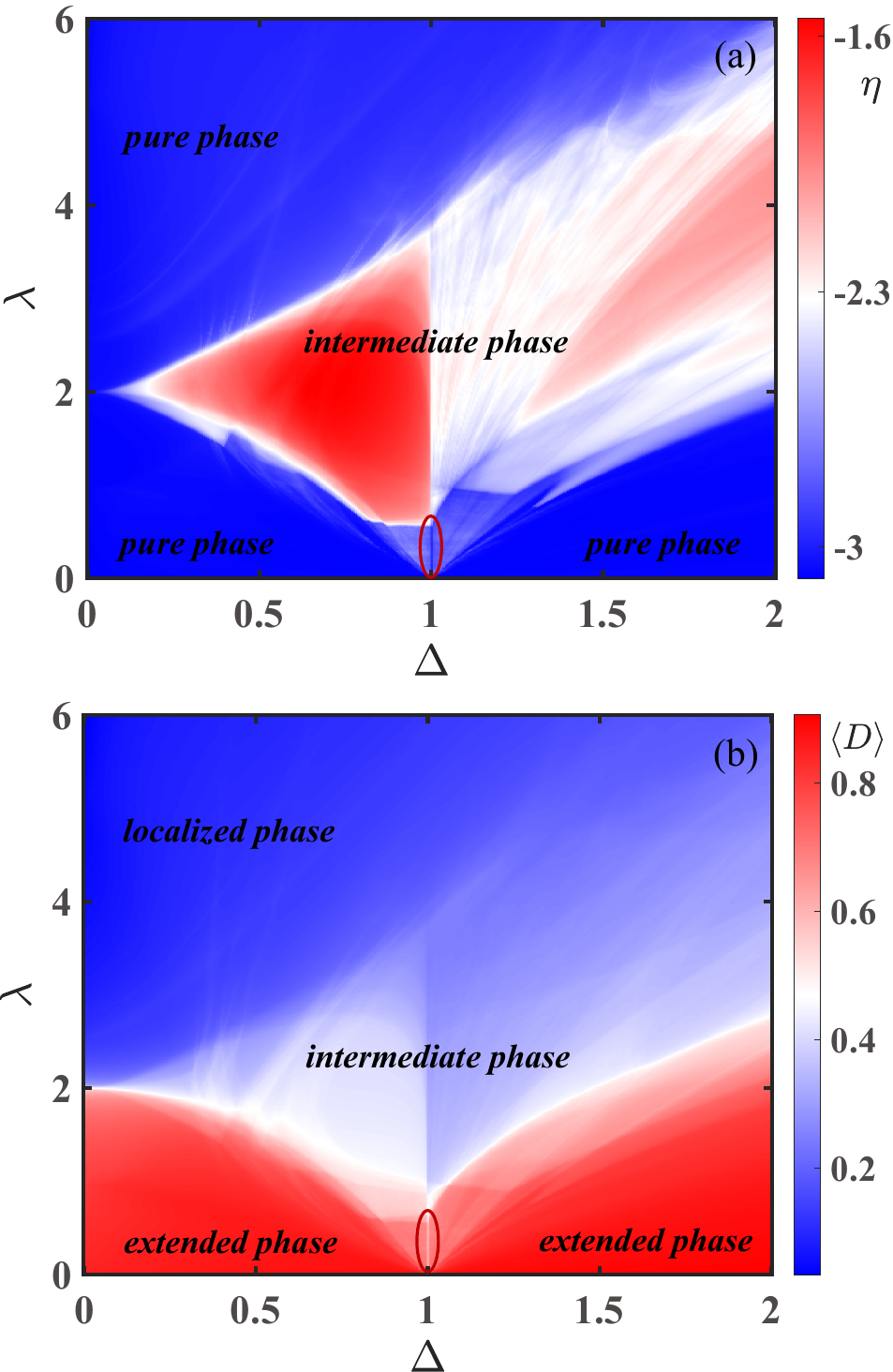}
 \caption{(Color online). Phase diagram that varies with ($\Delta,\lambda$) in terms of $\eta$ (a) and $\langle D \rangle$ (b). 
The pure phase (intermediate phase) is corresponding to the blue (red) region in (a). The pure phase is further distinguished into extended (deep red region) and localized phase (deep blue region) in (b).  
The critical phase is persists along $\Delta=1$ with $\lambda\lesssim0.57$, marked in red elliptic. Here, we set the parameter $\beta=0.01$ and $L=610$.}
 \label{fig.2}
\end{figure}

\begin{figure*}
\centering
 \includegraphics[width=17.5cm]{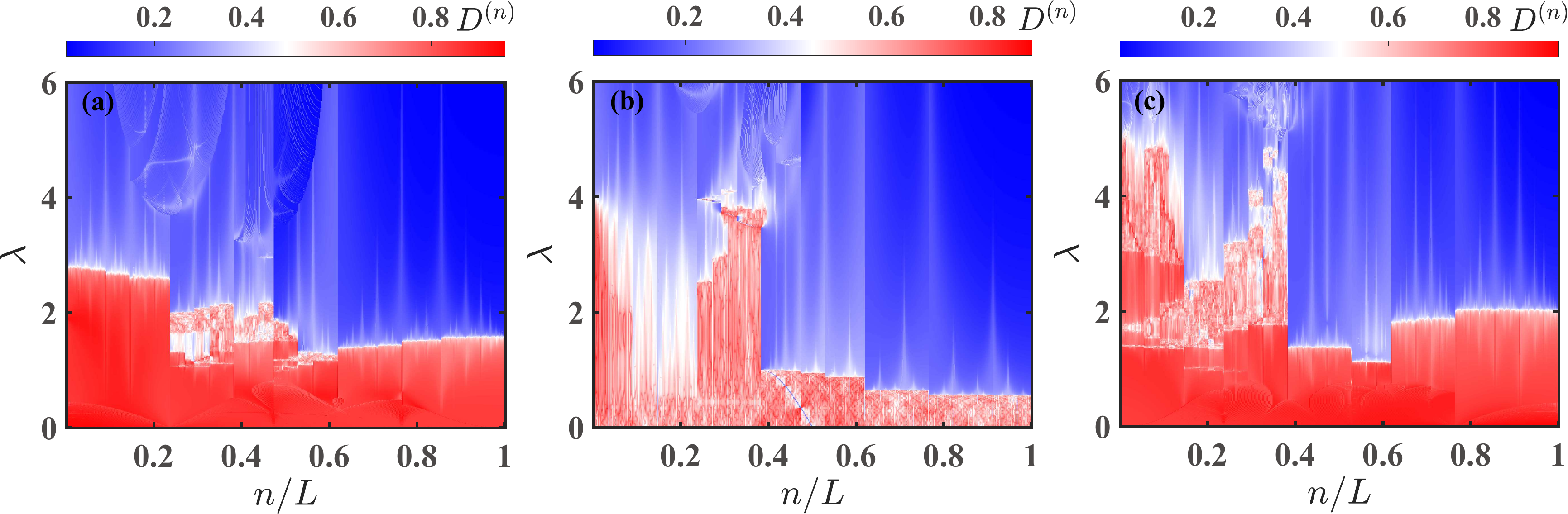}
 \caption{(Color online). (a)-(c) The fractal dimension $D^{(n)}$ versus $\lambda$, where $n$ denotes $n\rm{\mbox{-}th}$ eigenstate of BdG Hamiltonian. The parameter $\Delta$ is $0.5$ (a), $1$ (b) and $1.5$ (c), respectively. Other parameters are $\beta=0.01$ and $L=610$.}
 \label{fig.3}
\end{figure*}

\begin{figure}
\centering
 \includegraphics[width=8.5cm]{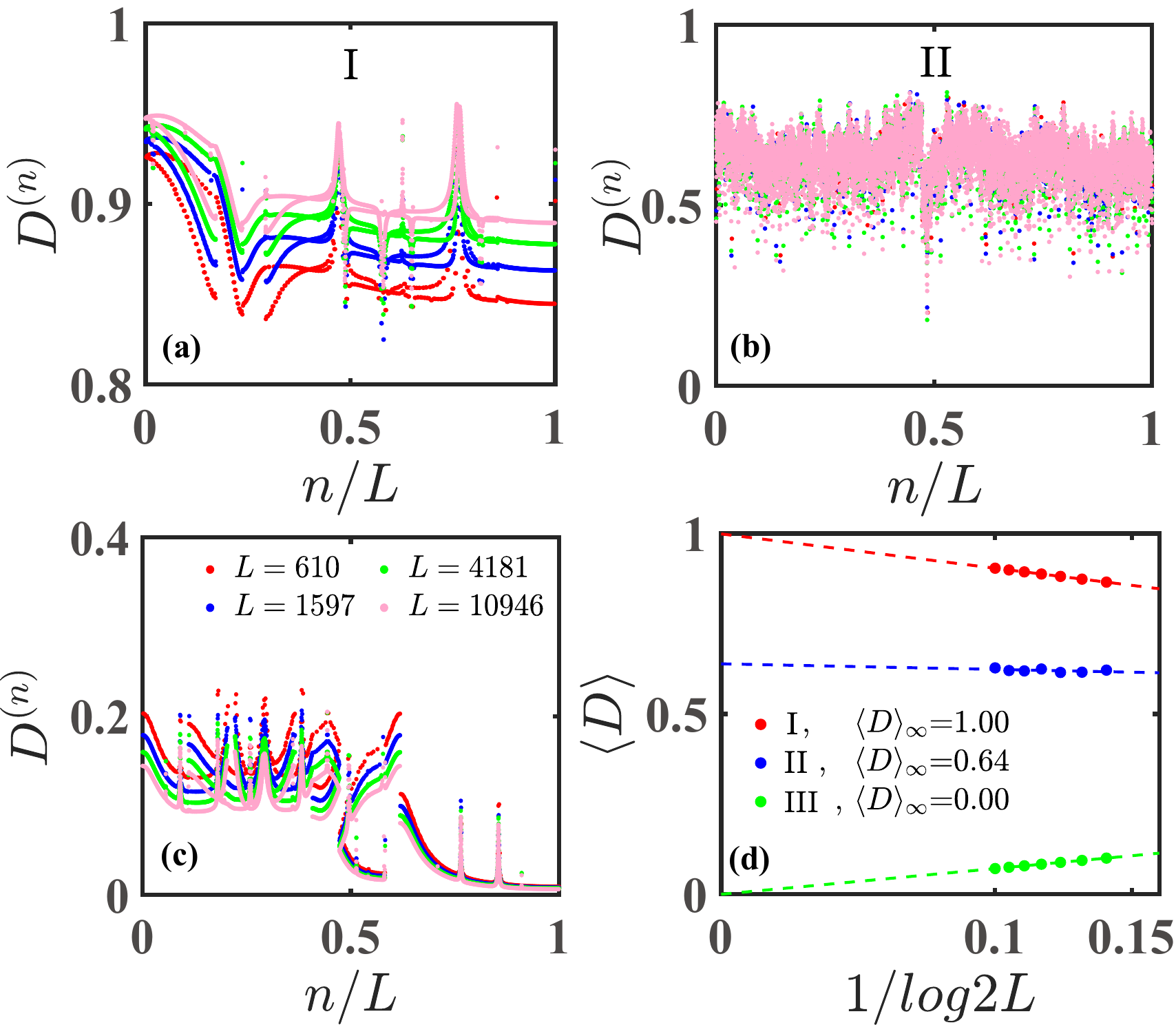}
 \caption{(Color online). (a)-(c) The fractal dimension $D^{(n)}$ for different $L$ at fixed 
$\Delta=0.5$ and $\lambda=0.2$ (a),
$\Delta=1$ and $\lambda=0.2$ (b),
$\Delta=0.5$ and $\lambda=5.8$ (c), where $n$ denotes $n\rm{\mbox{-}th}$ eigenstate. 
(d) Finite-size extrapolation of $\langle D \rangle$ as a function $1/\rm{log}(2L)$.}
 \label{fig.4}
\end{figure}

In this following, we discuss several physical quantities to characterize the nature of wave function.
Firstly, we introduce the inverse participation ratio $({\rm{IPR}})$ defined in eq. (\ref{6}) and the normalized participation ratio $({\rm{NPR}})$ defined in eq. (\ref{7}), which are utilized to differentiate among the extended, critical and localized states~\cite{intermediate.Shilpi.2023}.
\begin{align}
\label{6}
{\rm{IPR}}^{(n)}&=\sum_{m=1}^{L}[|u_{m}^{(n)}|^4+|v_{m}^{(n)}|^4], \\
\label{7}
{\rm{NPR}}^{(n)}&=[2L \times {\rm{IPR}}^{(n)}]^{-1}.
\end{align}
Where eq. (\ref{6}) satisfies ${\rm{IPR}}^{(n)}\sim1/(2L)^{\gamma_n}$~\cite{MobilityEdge12.2023}.
The index $n$ denotes the $n\rm{\mbox{-}th}$ eigenstate of BdG Hamiltonian and $m$ is the $m\rm{\mbox{-}th}$ element of that eigenstate.
For the $n\rm{\mbox{-}th}$ eigenstate, when ${\rm{IPR}}$ approaches $0$, it indicates a extended state and the corresponding $\gamma_n=1$.
Conversely, ${\rm{NPR}}$ approaches $0$ and $\gamma_n=0$ for a localized state.
If the state is critical, $\gamma_n\in(0,1)$.

For a large-size system, the fractal dimension $D^{(n)}$ is defined as follows
~\cite{MobilityEdge72.2019, MobilityEdge8.2020, MobilityEdge12.2023}:
\be 
\label{8}
D^{(n)}=-\lim_{L\rightarrow\infty}\frac{\log({\rm{IPR}}^{(n)})}{\log 2L}.
\ee
By analyzing the inverse participation ratio ${\rm{IPR}}^{(n)}$, we can easily infer that $D^{(n)}$ goes to $0$ ($1$) for the localized (extended) state and $D^{(n)}\in(0,1)$ for the critical state. 
Then the average fractal dimension $\langle D \rangle$ averaged over the BdG quasiparticle spectrum can capture the overall characteristics of the system and it is defined as:
\begin{align}
\label{9}
\langle D \rangle = \frac{1}{L} \sum_{n=1}^{L} D^{(n)}.    
\end{align}
The system exhibits phases that are either extended, where the average fractal dimension $\langle D \rangle$ approaches $1$, or localized, where $\langle D \rangle$ approaches $0$.
However, it cannot distinguish the critical phase from intermediate phase. 
It is necessary to compute $D^{(n)}$ for each eigenstate, if $D^{(n)}\in(0,1)$ for all the eigenstates, it suggests a critical phase. 
Furthermore, we define $\overline{D}$ averaged across a subset of eigenstates to capture the different states coexist in an intermediate phase. 

%
Next, we introduce $\eta$ which aids in distinguishing pure phases (extended or localized phase) from intermediate phase, defined as~\cite{MobilityEdge72.2019,
MobilityEdge8.2020, MobilityEdge12.2023},
\be 
\label{10}
\eta={\rm{log_{10}}}[\langle{\rm{IPR}}\rangle \times \langle{\rm{NPR}}\rangle],
\ee
where $\langle{\rm{IPR}}\rangle$ and $\langle{\rm{NPR}}\rangle$ are given by eq. (\ref{11}).  
For the extended phase, $\langle{\rm{IPR}}\rangle \rightarrow 0$ [$\langle{\rm{NPR}}\rangle \rightarrow$ finite].
Conversely, for the localized phase, $\langle{\rm{NPR}}\rangle \rightarrow 0$ [$\langle{\rm{IPR}}\rangle \rightarrow$ finite].
So we have 
$\langle{\rm{IPR}}\rangle \times \langle{\rm{NPR}}\rangle \sim 1/2L$ and
$\eta\lesssim-3$ in the pure phases, where $L=610$ in Fig.~\ref{fig.2}. 
For the intermediate phase, both of $\langle{\rm{IPR}}\rangle$ and $\langle{\rm{NPR}}\rangle$ keep finite and $-3<\eta\lesssim-1$.

\begin{align}
\label{11}
&\langle \mathrm{IPR} \rangle = \frac{1}{L} \sum_{n=1}^{L} \mathrm{IPR}_n, 
&\langle \mathrm{NPR} \rangle = \frac{1}{L} \sum_{n=1}^{L} \mathrm{NPR}_n.
\end{align}

\section{phase diagram for Small $\beta$}
\label{sec3}
As mentioned above, the potential $V_i(\beta)$ simplifies to 
${\rm{cos}}(2\pi \alpha i+\theta)-{\rm{cos}}(\pi \alpha)$ in small $\beta$ limit.
The system enters into the localized phase when $\lambda$ exceeds a critical threshold due to the existence of quasiperiodic potential, i.e., $\alpha$ is incommensurate (see Appendix.~\ref{secs1}).
It differs from the previous study which exclusively considered a cosine potential without the constant on-site chemical potential shift~\cite{pair.Cai.2013}. 
In order to substantiate this distinction, we show the phase diagram where variable $\eta$ and fractal dimension $\langle D \rangle$ versus ($\lambda$, $\Delta$) in Fig.~\ref{fig.2}.
This diagram features two distinct regions: the pure phases (depicted in blue) and various intermediate phases (depicted in red) separated by $\eta$ in Fig.~\ref{fig.2}\hyperref[fig.2]{(a)}, which is the key point in our paper.
The pure phases are further distinguished into extended phases (deep red region) and localized phases (deep blue region) by $\langle D \rangle$ in Fig.~\ref{fig.2}\hyperref[fig.2]{(b)}.

In order to have a complete insight into the phase diagram, we calculate the fractal dimension $D^{(n)}$ where $n$ denotes $n\rm{\mbox{-}th}$ eigenstate of BdG Hamiltonian  versus the potential strength $\lambda$ for different $\Delta=0.5$, $1$ and $1.5$, as illustrated in Fig.~\ref{fig.3}. 
It is interesting that the critical phase is confined to a narrow line where $\Delta=1$ and $\lambda\lesssim0.57$.
Therefore, the complete phase diagram includes three pure phases (extended, localized and critical phase) and many types of intermediate phases. 
A more detailed discussion of these phases will be provided in subsequent sections.

\subsection{Pure phases}
\label{sec31}
Fig.~\ref{fig.3}\hyperref[fig.3]{} shows the system is in the extended (critical) phase where all the states are extended (critical) when $\Delta\neq1$ ($\Delta=1$), with weak potential strength $\lambda$.
To provide more precise numerical evidences, we further calculate the $D^{(n)}$ for various lattice sizes $L$ at selected values of $\lambda$, the results are displayed in Fig.~\ref{fig.4}\hyperref[fig.4]{(a)-(b)}. And the finite-size extrapolation of $\langle D \rangle$ averaged over the quasiparticle spectrum is shown in Fig.~\ref{fig.4}\hyperref[fig.4]{(d)}. 
Take $\Delta=0.5$ and $\lambda=0.2$ for example, Fig.~\ref{fig.4}\hyperref[fig.4]{(a)} shows the $D^{(n)}$ for all the states increases with $L$ and $\langle D \rangle$ approaches $1$ in the thermodynamic limit, 
indicating the system is in the extended phase.
Additionally, when $\Delta=1$, the $D^{(n)}$ fluctuates from $0$ and $1$, independent on $L$, indicating all the states are critical, as shown in Fig.~\ref{fig.4}\hyperref[fig.4]{(b)}.
When the potential strength $\lambda$ is strong enough, such as $\lambda=5.8$ for $\Delta=0.5$, the system goes to a localized phase where $D^{(n)}$ for all the states decreases with $L$ and $\langle D \rangle$ tends to $0$ in the thermodynamic limit, as shown in Fig.~\ref{fig.4}\hyperref[fig.4]{(c)}.
Therefore, the system exhibits three distinct pure phases, including extended, critical and localized phase.

\begin{figure}
\centering
 \includegraphics[width=8.5cm]{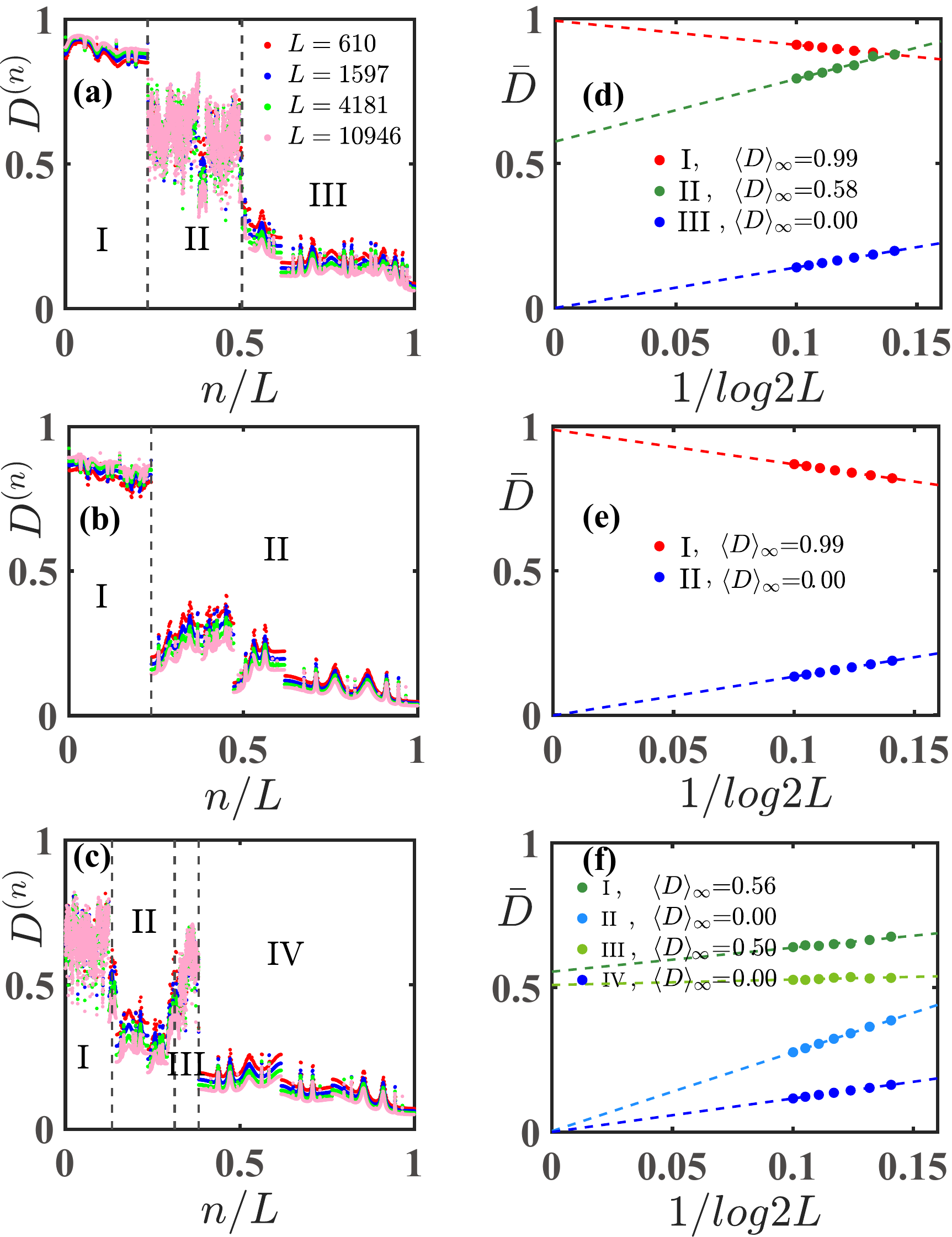}
 \caption{(Color online). (a)-(c) The fractal dimension $D^{(n)}$ for different $L$ at fixed 
$\Delta=0.5$ and $\lambda=1.8$ (a),
$\Delta=0.5$ and $\lambda=2.3$ (b),
$\Delta=1.5$ and $\lambda=3.5$ (c), where $n$ denotes $n\rm{\mbox{-}th}$ eigenstate. 
(d)-(f) Finite-size extrapolation of $\langle D \rangle$ as a function $1/\rm{log}(2L)$ averaged over the different state zones in (a)-(c).}
 \label{fig.5}
\end{figure}

\begin{figure}
\centering
 \includegraphics[width=8.6cm]{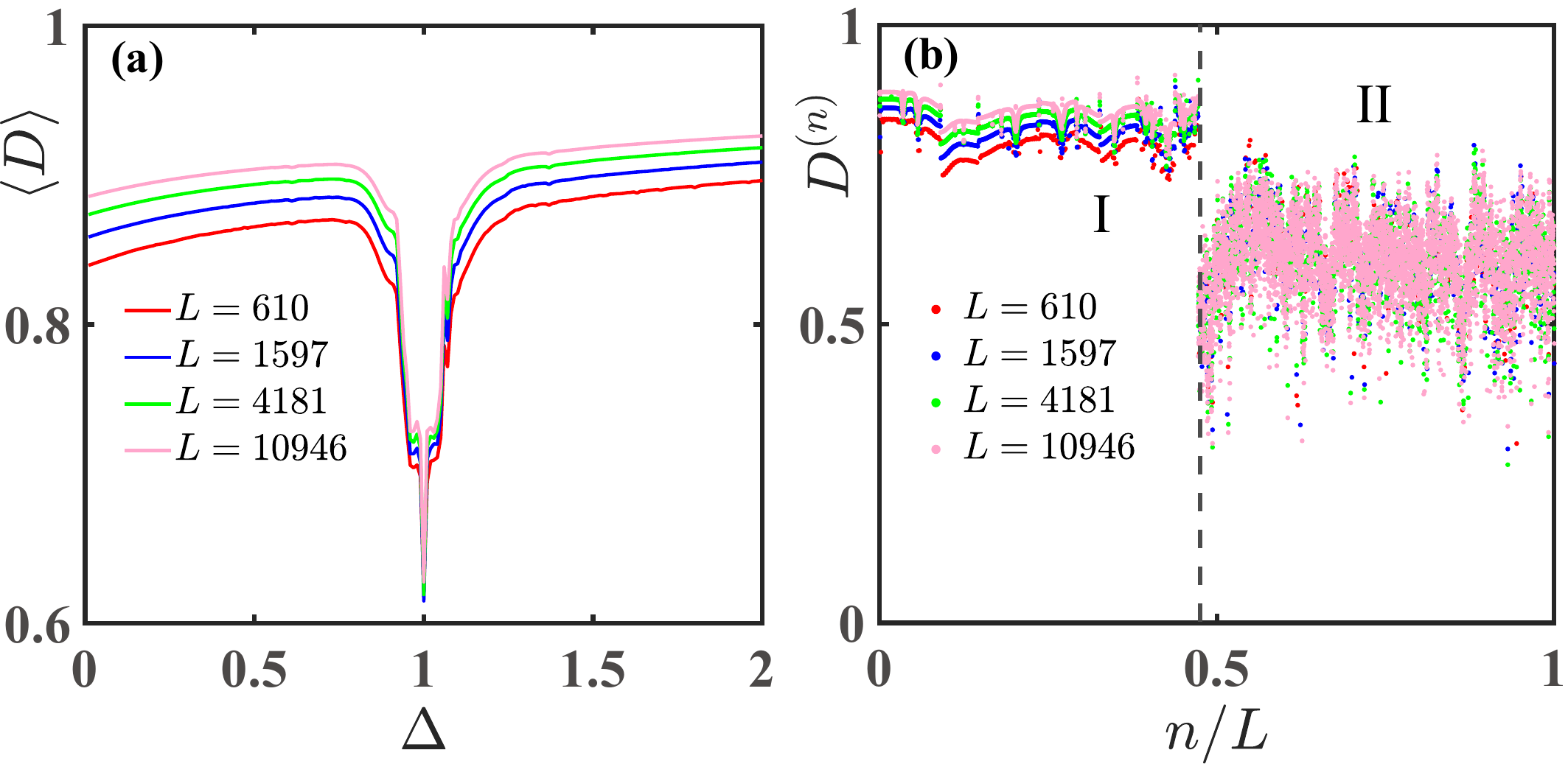}
 \caption{(Color online). (a) The fractal dimension $\langle D \rangle$ averaged over all the BdG quasiparticle spectrum versus $\Delta$.
(b) The fractal dimension $D^{(n)}$ for different $L$ at $\Delta=1$, where $n$ denotes $n\rm{\mbox{-}th}$ eigenstate. The parameter $\lambda=0.2$.}
 \label{fig.6}
\end{figure}

\subsection{Intermediate phases}
\label{sec32}
One can observe that the system undergoes various intermediate phases before transitioning into the localized phase as shown in Fig.~\ref{fig.3}\hyperref[fig.3].
When $\Delta=0.5$ and $\lambda=1.8$ as shown in Fig.~\ref{fig.5}\hyperref[fig.5]{(a)} \hyperref[fig.5]{(d)}, $D^{(n)}$ corresponding to the low energy states in zone $\rm{I}$ increases with $L$,
and the finite-size extrapolation of $\overline{D}$ averaged over the zone $\rm{I}$
goes to $1$, indicating all the states in zone $\rm{I}$ are extended.
While $D^{(n)}$ corresponding to the high energy states in zone $\rm{III}$ decrease with $L$,
and the finite-size extrapolation of $\overline{D}$ averaged over the zone $\rm{III}$ goes to $0$, indicating all the states in zone $\rm{III}$ are localized.
In contrast, $D^{(n)}$ for the states in zone $\rm{II}$ fluctuates around $0.6$, almost independent of $L$,
and the finite-size extrapolation of $\overline{D}$ averaged over the zone $\rm{II}$ approaches a finite value between $0$ and $1$, indicating all the states in zone $\rm{II}$ are critical.
Hence the system exhibits an intermediate phase with coexisting localized, extended, and critical states. 
These states are separated by the two types of anomalous mobility edge separating  extended or localized from critical states.
When $\lambda$ is slightly increased (i.e., $\lambda=2.3$) shown in Fig.~\ref{fig.5}\hyperref[fig.5]{(b)(e)}, we identify another intermediate phase with coexisting localized and extended states where $\overline{D}$ goes to $1$ ($\rm{I}$) and $0$ ($\rm{II}$) in the thermodynamic limit, respectively. 
This intermediate phase exhibits a traditional mobility edge separating the extended and localized zones.
When the $\Delta=1.5$ and $\lambda=3.5$ shown in Fig.~\ref{fig.5}\hyperref[fig.5]{(c)(f)}, 
$D^{(n)}$ for states in zones $\rm{II}$ and $\rm{IV}$ decrease with $L$ and its average value $\overline{D}$ goes to $0$, indicating they are localized. 
Conversely, $D^{(n)}$ fluctuates around $0.56$ and $0.5$ for states in zones $\rm{I}$ and $\rm{III}$ , indicating the states are critical. Therefore, the system has an intermediate phase with coexisting localized and critical states.
These states in different zones are separated by an anomalous mobility edge.  

%
The system is known to exhibit a critical phase when $\Delta=1$ and $\lambda\lesssim0.57$ (see Sec.~\ref{sec31}).  Additionally, a distinct intermediate phase emerges when $\Delta$ is slightly deviating from $1$.
Now take $\lambda=0.2$ for example, 
Fig.~\ref{fig.6}\hyperref[fig.6]{(a)} shows that the $\langle D \rangle$ varies smoothly versus $\Delta$ when $\Delta\lesssim0.92$.
A notable decrease in $\langle D \rangle$ is first observed when the system goes into the intermediate phase with coexisting extended ($\rm{I}$) and critical states ($\rm{II}$), as shown in Fig.~\ref{fig.6}\hyperref[fig.6]{(b)}. 
Subsequently, a second notable decline occurs when the system enters the critical phase with $\Delta=1$ [see Fig.~\ref{fig.4}\hyperref[fig.4]{(b)}]. 
The phenomenon is easily understand by the ultimate value of $D^{(n)}$ is less than $1$ in the thermodynamic limit for the critical states.
Consequently, the system exhibits four distinct intermediate phases: the first one with coexisting extended and localized states; the second one with coexisting extended and critical states; the third one with coexisting localized and critical states; and the fourth one with coexisting extended, critical, and localized states.

\begin{figure}
\centering
 \includegraphics[width=9.2cm]{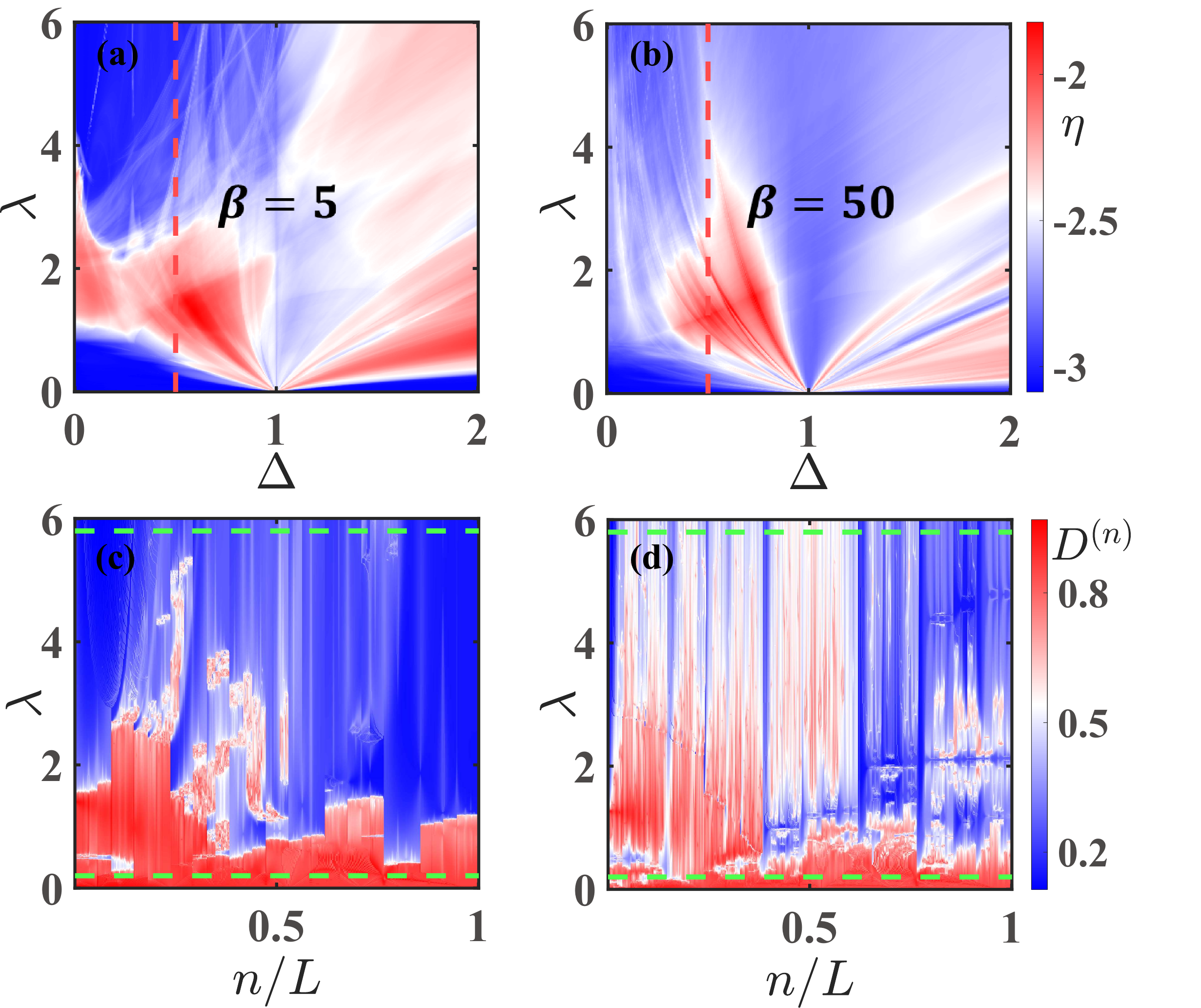}
 \caption{(Color online). (a)-(b) Phase diagram of variable $\eta$ versus ($\Delta,\lambda$). The red dashed line corresponds to $\Delta=0.5$.
(c)-(d) The fractal dimension $D^{(n)}$ versus $\lambda$, where $n$ denotes $n\rm{\mbox{-}th}$ eigenstate of BdG Hamiltonian when $\Delta=0.5$. The green dashed lines corresponds to $\lambda=0.2$ and $5.8$, respectively. Here, the parameter $\beta=5$ (a)(c), $50$ (b)(d) and $L=610$ .}
 \label{fig.7}
\end{figure}

\begin{figure*}
\centering
 \includegraphics[width=17.5cm]{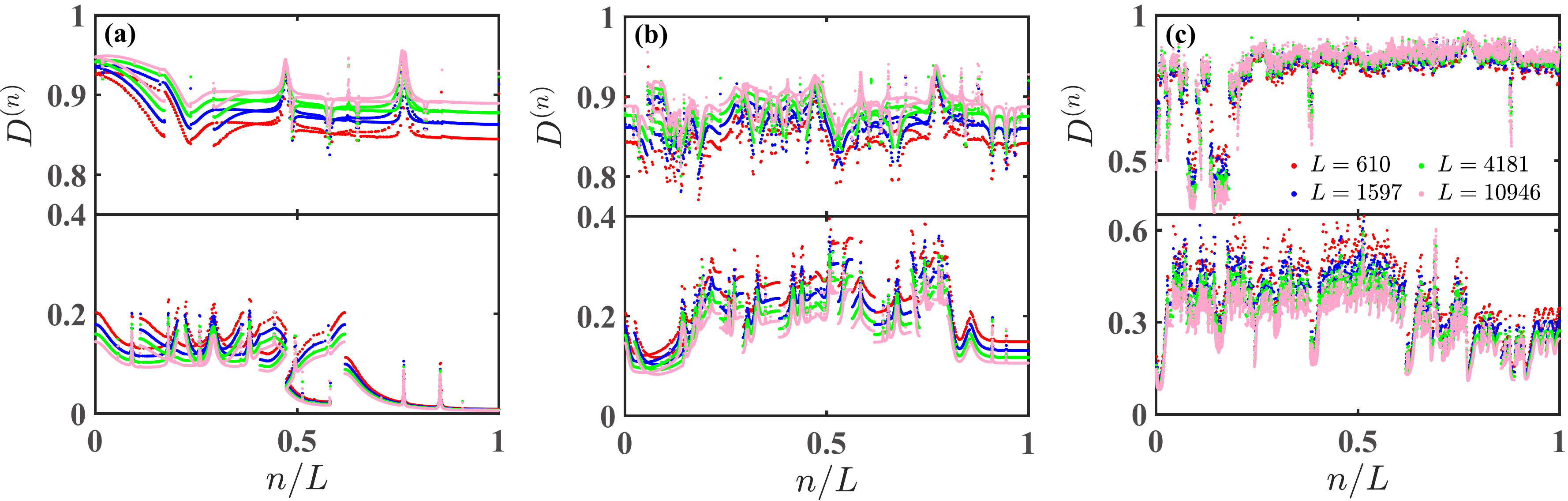}
 \caption{(Color online). The fractal dimension $D^{(n)}$ for different $L$ at fixed $\beta=0.01$ (a), $\beta=5$ (b), $\Delta=50$ (c), where $n$ denotes $n\rm{\mbox{-}th}$ eigenstate. Other parameters $\Delta=0.5$ and $\lambda=0.2$ ($5.8$) for the upper (lower) panel of each figure.}
 \label{fig.8}
\end{figure*}

\section{phase diagram with increasing $\beta$}
\label{sec4}
In order to investigate the effect of increasing $\beta$,
we begin by analyzing the phase diagram where variable $\eta$ versus ($\lambda$, $\Delta$) at fixed 
$\beta=5$ and $\beta=50$, as shown in Fig.~\ref{fig.7}\hyperref[fig.7]{(a)} and \hyperref[fig.7]{(b)}. 
In contrast to the phase diagram shown in Fig.~\ref{fig.2}\hyperref[fig.2]{(a)}, we found that the zones for pure phases such as extended phase and localized phase are significantly diminished with increasing $\beta$. 
We plot the fractal dimension $D^{(n)}$ versus $\lambda$ at fixed $\Delta=0.5$ in Fig.~\ref{fig.7}\hyperref[fig.7]{(c)-(d)}.
Fig.~\ref{fig.7}\hyperref[fig.7]{(c)} shows that an extensive number of extended states are replaced by the critical states or localized states when the strength of potential is weak, which differs markedly from the scenario presented in Fig.~\ref{fig.3}\hyperref[fig.3]{(a)}.
As the strength of potential $\lambda$ increases,
the system exhibits various intermediates phases, such as one comprising both localized and critical states, and another with coexisting extended, critical and localized states.
What is more, the system goes to the localized phase when the strength of potential is further increased. Hence, the phase diagram does not have essential changes when $\beta=5$.
However, the pure phases (extended or localized phase) almost disappear and more critical states emerge when further increasing $\beta$, as shown in Fig.~\ref{fig.7}\hyperref[fig.7]{(d)}.

To illustrate the impact of $\beta$ on the system, we examine the fractal dimension $D^{(n)}$ for different $L$ at a fixed weak $\lambda=0.2$ and a strong $\lambda=5.8$, with $\beta$ ranging from small to large, as shown in the upper and lower panels of Fig.~\ref{fig.8}, respectively.
The $D^{(n)}$ goes to $1$ ($0$) with the increase of $L$ when $\beta=0.01$, representing a completely extended (localized) phase, as shown in Fig.~\ref{fig.8}\hyperref[fig.8]{(a)}. 
At $\beta=5$, $D^{(n)}$ exhibits minor fluctuations for a limited number of states, suggesting a slightly deviation from pure phase, yet without significantly altering its essence, as shown in Fig.~\ref{fig.8}\hyperref[fig.8]{(b)}.
However, this fluctuation becomes pronounced at higher $\beta$ values. 
Fig.~\ref{fig.8}\hyperref[fig.8]{(c)} reveals that $D^{(n)}$ fluctuates between $0$ and $1$ for more states, indicating the critical states.
Hence the system has an intermediate phase with coexisting more critical states and less localized states.
One can infer that the localized states will be completely replaced by the critical states with further increments in $\beta$, which is confirmed in Appendix~\ref{secs21}. Additionally, we also explore the fate of the critical phase for large $\lambda$, the details are shown in the Appendix~\ref{secs22}).

\begin{figure*}
\centering
 \includegraphics[width=17.5cm]{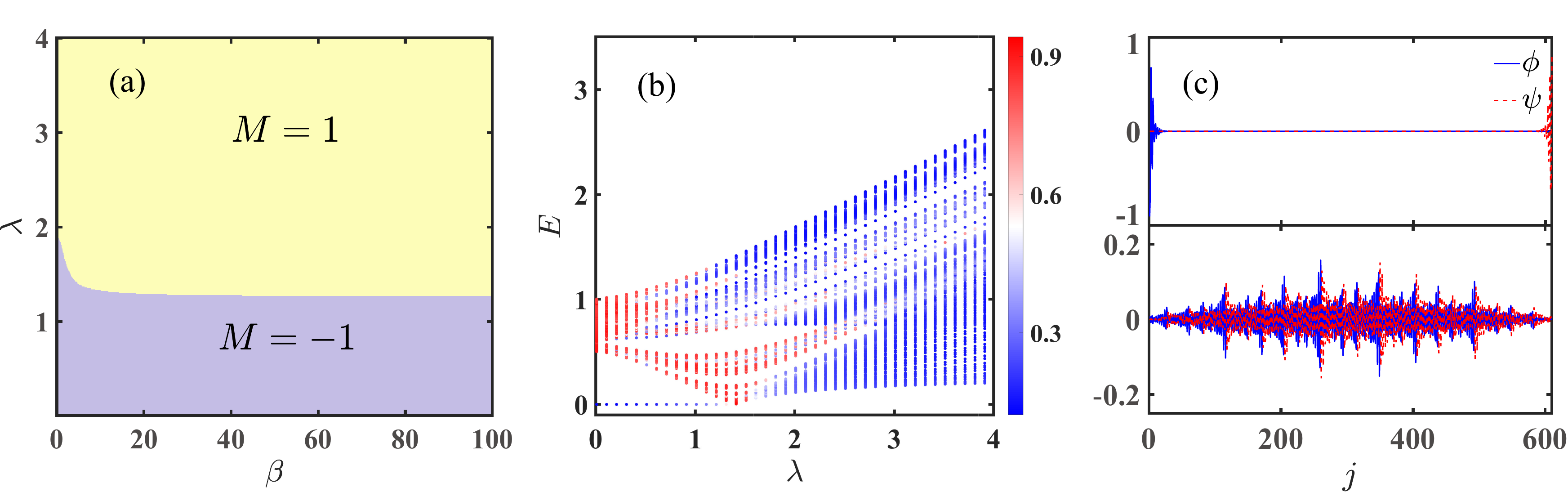}
 \caption{(Color online). (a) Phase diagram that varies with $(\beta,\lambda)$ in terms of $Z_2$ topological invariant $M$. (b) Energy spectrum versus $\lambda$ for $L=610$, The color coding correlates to the fractal dimension $D^{(n)}$. (c) The spatial distributions of $\phi$ and $\psi$ for the lowest excitation with $V=1.0$ (upper panel) and $V=1.5$ (lower panel) where $j$ is $j\rm{\mbox{-}th}$ site. The parameter $\beta=5$ for (b)-(c). Open boundary condition.}
 \label{fig.9}
\end{figure*}

\section{topological properity}
\label{sec5}
In this section, we study the topological properties of system using $Z_2$ topological invariant $M$, which emerged as a frontier in understanding the properties of materials that exhibit topological superconductivity\cite{MZM_2001_Kitaev}.
We employ the numerical method referenced in \cite{MZM_2011} to calculate $Z_2$ topological invariant in the open boundary condition (OBC). 
The eq. (\ref{52}) can be simplified into the following form with zero-mode states $\epsilon_1=0$ ($\mu=1$):
\begin{align}
\label{s1}
&(A+B)\phi_1 =0, 
&(A-B)\psi_1 =0,
\end{align}

In the transfer matrix form, $\phi_1$ and $\psi_1$ can be rewritten as:
\begin{align}
\label{s2}
[\phi_1(i-1),\phi_1(i)]^{T}=F_i[\phi_1(i),\phi_1(i+1)]^{T},\nonumber \\
[\psi_1(i+1),\psi_1(i)]^{T}=F_i[\psi_1(i),\psi_1(i-1)]^{T},
\end{align}
where
\begin{align}
\label{s2}
F_i=\left(\begin{array}{cc}
\frac{\lambda V_i(\beta)}{\Delta+J} & \frac{\Delta-J}{\Delta+J}\\
1 & 0
\end{array}\right).
\end{align}

The total transfer matrix for system size $L$ is defined as $F=F_L F_{L-1}\cdots F_2F_1$. The $Z_2$ topological invariant $M$ is determined as $M=\rm{sgn}(ln(|\lambda_2|))$, where $\lambda_1$, $\lambda_2$ are eigenvalues of $F$ with the assumption $|\lambda_1|<|\lambda_2|$.
The topological phase diagram as a function of $(\beta,\lambda)$ is illustrated in Fig.~\ref{fig.9}\hyperref[fig.9]{(a)}, where $M=-1 (1)$ corresponds to the topological non-trivial (trivial) phase. 
We observe that the topological phase transition point is less than $3$ ~\cite{pair.Cai.2013} when $\beta\rightarrow0$, which emphasizes once again for the discrepancy with QP potential without constant on-site chemical potential shift.
The topological phase boundary decay slightly at the onset but quickly reach saturation with increasing $\beta$, indicating the topological nature does not change in the process of AA limit tending to the Fibonacci limit. 
%
Additionally, the topological phase transition is accompanied by the closure and subsequent reopening of the energy gap, as shown in Fig.~\ref{fig.9}\hyperref[fig.9]{(b)}. The gap narrows when $\lambda$ is small and may close completely when $\lambda$ intersects in the topological phase boundary, then the gap reopens as $\lambda$ further increases.  
%
To intuitively understand the behavior of the edge states, we present the lowest excitation state of $\phi$ and $\psi$ in Fig.~\ref{fig.9}\hyperref[fig.9]{(c)}. 
When $\lambda=1.0$, the distribution of zero-mode state $\phi$ ($\psi$) locates at left (right) end with a narrow spread, which is consistent with topological non-trivial phases, as shown in the upper panel of Fig.~\ref{fig.9}\hyperref[fig.9]{(c)}. When $\lambda=1.5$ within the trivial phase (see the lower panel), the $\phi$ and $\psi$ extended throughout the bulk, meaning the loss of majorana  edge states.

\section{Conclusion and outlook}
\label{sec6}
In summary, our research delineates the various quantum phases and the topological properities emerged in the IAAF model with p-wave SC pairing terms. 
For the former, this model exhibits modifiable phase diagrams through the tunable parameter $\beta$. 
For small values of $\beta$, this model can be reduced to the generalized AA model up to a constant on-site chemical potential shift. 
The system is always in the pure phases when the strength of potential is weak (extended or critical phase) or strong (localized phase) enough.
What is more,
it is interesting that the system has many types of intermediate phases when the strength of potential is moderate.
For instance, one can observe an intermediate phase where extended and localized states coexist, as well as phases where extended and critical states, or localized and critical states, are present concurrently. 
Also, the coexistence of extended, critical and localized states. 
These coexisting states are separated by different type of mobility edges.
As $\beta$ increase, the pure phases (extended or localized phase) will gradually diminish, and the system becomes critical in the Fibonacci limit.
For the latter, we observe the system transitions from topologically nontrivial to trivial phase via increasing $\lambda$.
The existence of the majorana edge states and the closure and reopening of the energy gap demonstrate the Z2 topological nature.

Experimentally, we expect that the proposed model can be realized in current superconducting circuit quantum simulator\cite{Exp1.2019,Exp2.2019,Exp3.2019}, where the nearest neighboring pairing can be realized as a consequence of coherent two-photon
driving. 
Even though in such a system, the particle is boson instead of fermion, one can tune the on-site repulsive interaction between the bosons to make them become hard-core. For a 1D system, a hard-core boson model is equivalent to a spinless fermion model even in the presence of nearest neighboring pairing. In such a synthetic quantum system, the parameters are highly tunable, which allows us to access the parameter regime studied in this paper.

This work unveils a quantum model that exhibits many types of intermediate phases, thereby enriching the understanding of mobility edges. A natural question is that whether these intermediate phases are robust when interactions are introduced.
Additionally, investigating the dynamic properties that arise from the various phases may be another intriguing question. 
Besides,  
the one-dimensional (1D) p-wave superconducting paired fermion model can be mapped onto the transverse XY model via the Jordan-Wigner transformation \cite{XY.FDS.1995, XY.YAP.1996}. Our research casts a new light on the study of analogous phenomena related to localization in low-dimensional quasi-periodic spin systems.

\section{Acknowledgments}
\label{sec7}
We thank Zi Cai for useful discussions. This work is supported by the National Key Research and Development Program of China 
(Grant No. 2020YFA0309000), NSFC of China (Grant No.12174251), the Natural Science Foundation of Shanghai (Grant No.22ZR142830), and the Shanghai Municipal Science and Technology Major Project (Grant No.
2019SHZDZX01).

\FloatBarrier
\appendix

\begin{figure}
\centering
 \includegraphics[width=8.5cm]{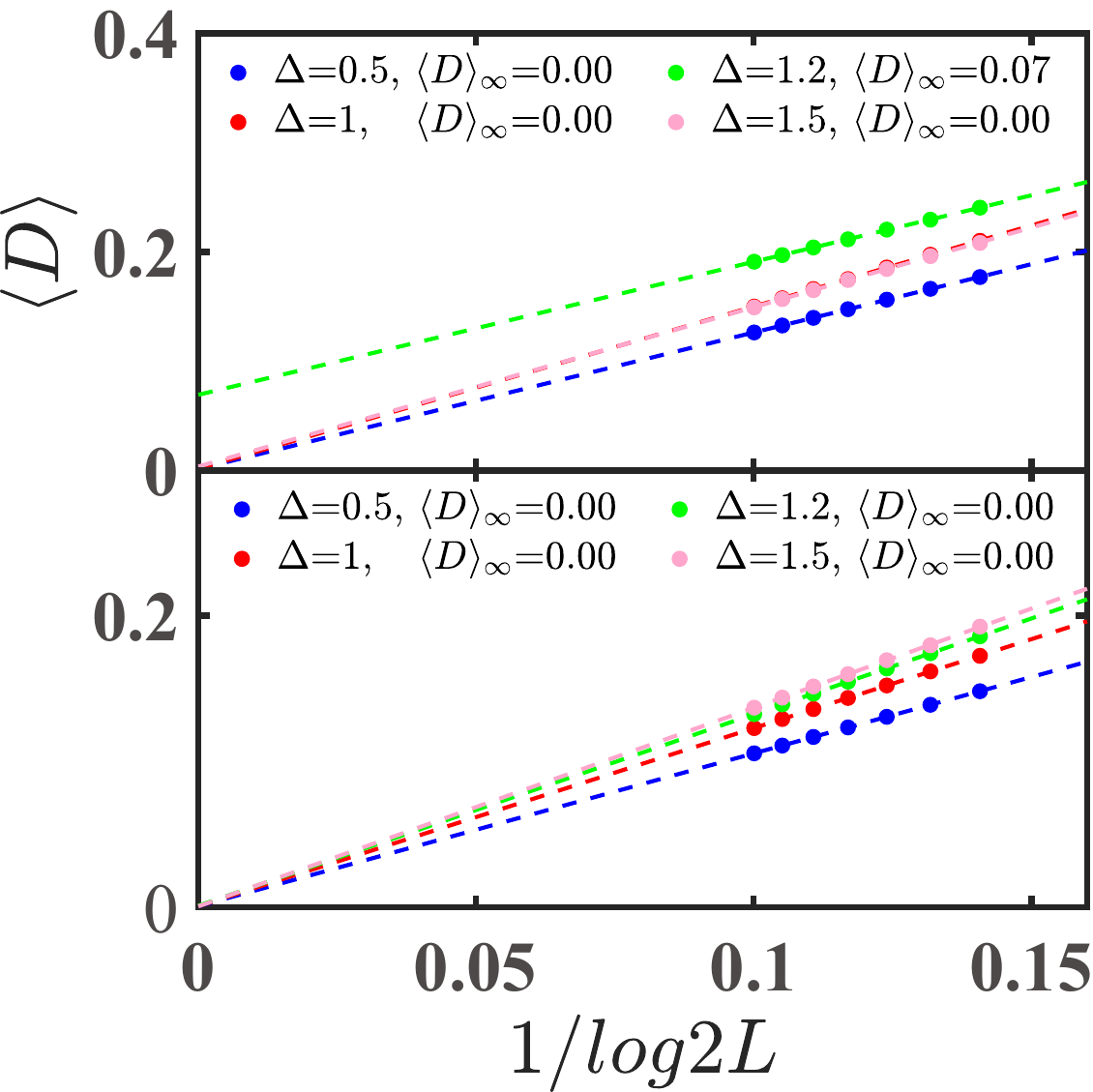}
 \caption{(Color online). Finite-size extrapolation of $\langle D \rangle$ as a function $1/\rm{log}(2L)$, where $\lambda=2\Delta+2$ (upper panel) and $\lambda=2\Delta+2.5$ (lower panel).}
 \label{fig.s1}
\end{figure}

\begin{figure}[tb]
\centering
 \includegraphics[width=8.0cm]{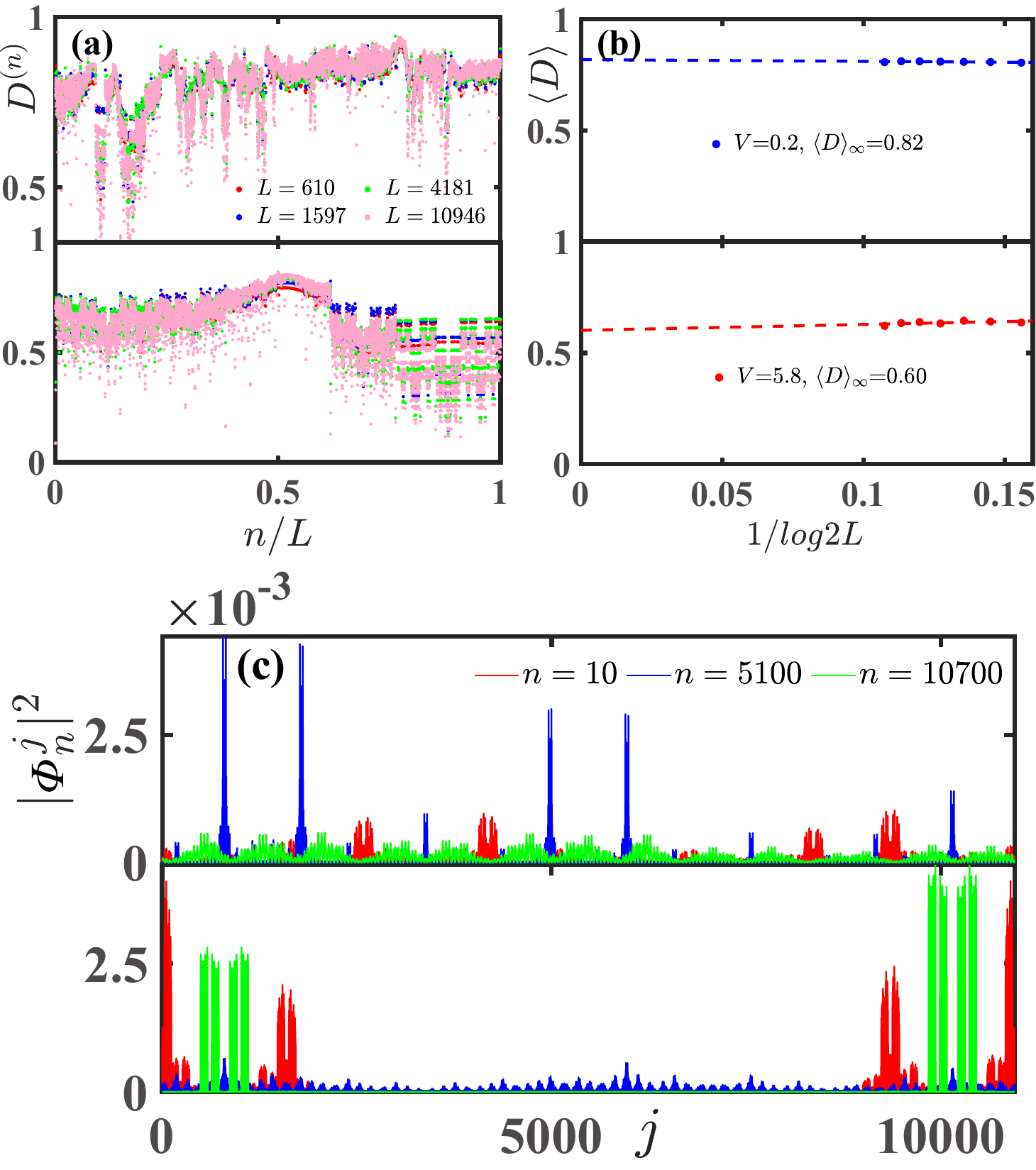}
 \caption{(Color online). (a) The fractal dimension $D^{(n)}$ for different $L$, where $n$ denotes $n\rm{\mbox{-}th}$ eigenstate. (b) Finite-size extrapolation of $\langle D \rangle$ as a function $1/\rm{log}(2L)$. (c) The probability distribution of selected ($n\rm{\mbox{-}th}$) eigenstate for  $L=10946$, where index $j$ denotes $j\rm{\mbox{-}th}$ site. The parameter $\lambda=0.2$ ($5.8$) for the upper (lower) panel of each figure, $\Delta=0.5$, $\beta=10^4$.}
 \label{fig.s21}
\end{figure}

\section{Breakdown of phase boundary in the small $\beta$ limit}
\label{secs1}
Ref.~\cite{ pair.Cai.2013} deduced the delocalization-localization transition at $\lambda_c=2(J+\Delta)$ for the Hamiltonian with cosine QP potential.
In the main text, $V_i(\beta)$ simplifies to ${\rm{cos}}(2\pi \alpha i+\theta)-{\rm{cos}}(\pi \alpha)$, reflecting the cosine QP potential with a constant on-site chemical potential shift in the small $\beta$ limit.
Thus Anderson's theorem is also not applicable here due to the presence of QP potentials.
While the phase boundaries $\lambda_c=2(J+\Delta)$ demarcating the different phases have become distorted under the effect of ${\rm{cos}}(\pi \alpha)$, as shown in Fig.~\ref{fig.s1}. 
The upper panel of Fig.~\ref{fig.s1} reveals that the finite-size extrapolation of $\langle D \rangle$ cannot goes to zero for certain values of $\Delta$ (e.g. $\Delta=1.2$), indicating the inability to transition into the localized phase. However, the system can be localized for larger $\lambda$ (e.g. $2\Delta+2.5$), as shown in the lower panel of Fig.~\ref{fig.s1}.
This suggests that, within the parameter space considered, the system experiences a localized phase for $\lambda$ values in the vicinity of $2(J+\Delta)$, indicating a slight distortion of the phase boundary.

\begin{figure*}
\centering
 \includegraphics[width=17.5cm]{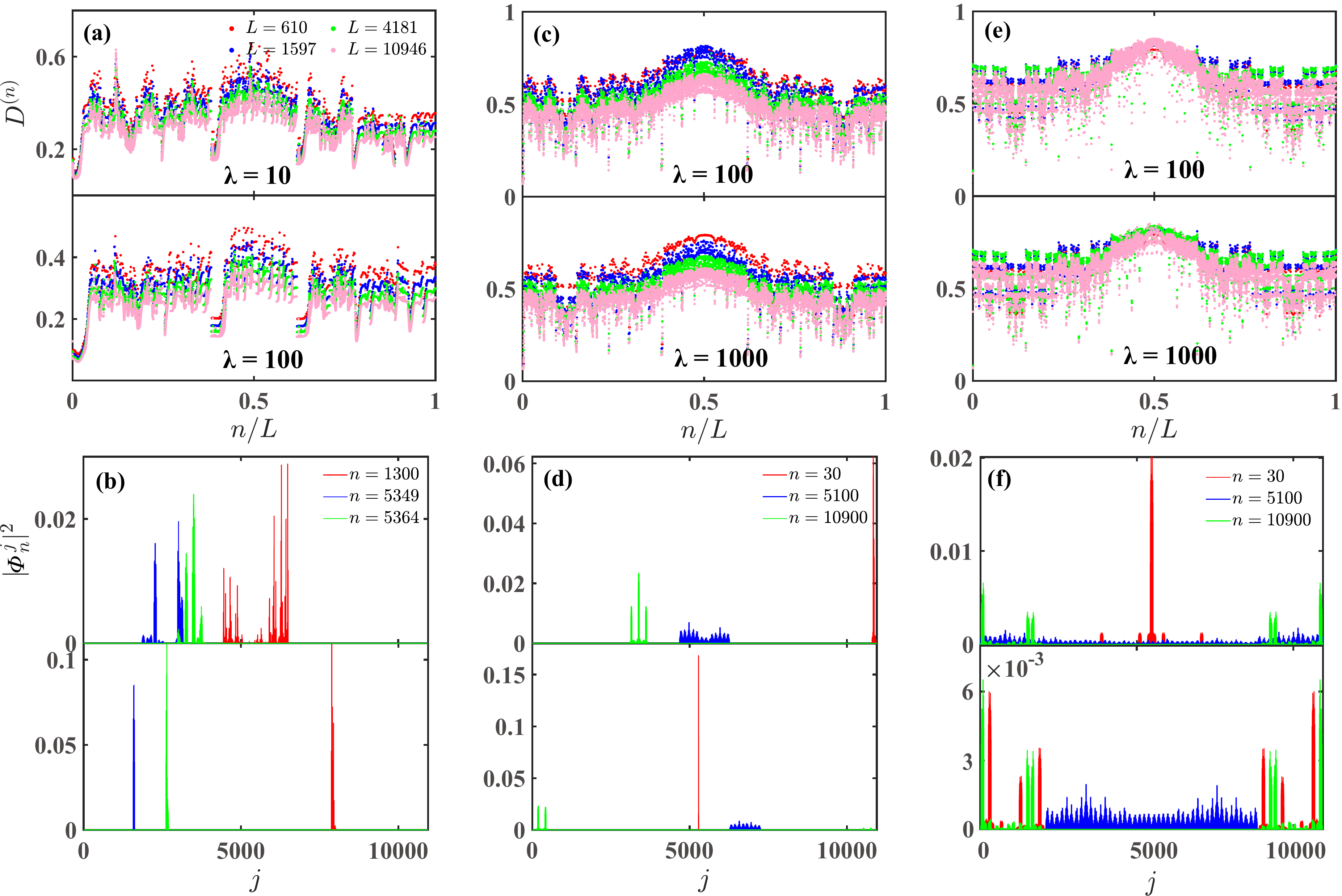}
 \caption{(Color online). The fractal dimension $D^{(n)}$ for different $L$ , where $n$ denotes the $n\rm{\mbox{-}th}$ eigenstate and $\beta=50$ (a), $\beta=10^3$ (c), $\beta=10^4$ (e). The parameters $\lambda=10$ ($100$) for the upper (lower) panel of (a), $\lambda=100$ ($1000$) for the upper (lower) panel of (c)(e). (b)(d)(f) The probability distributions of selected ($n\rm{\mbox{-}th}$) eigenstate for $L=10946$, where $j$ denotes the $j\rm{\mbox{-}th}$ lattice site. Other parameters are same with (a)(c)(e), respectively.}
 \label{fig.s22}
\end{figure*}

\section{Fate of the critical phase with large $\beta$}
\label{secs2}
\subsection{The effect of small $\lambda$}
\label{secs21}
In the main text, we find the pure phases (extended or localized phase) almost disappear and more critical states emerge when further increasing $\beta$. 
So we infer that system is critical when the $\beta$ is further increased (e.g. $\beta=10^4$), which is confirmed in Fig.~\ref{fig.s21}.
The distribution of $D^{(n)}$ for all the eigenstates fluctuates between $0$ and $1$, and the finite-size extrapolation of $\langle D \rangle$ averaged over the quasiparticle spectrum converges to a finite value for both of $\lambda=0.2$ and $\lambda=5.8$, as depicted in Fig.~\ref{fig.s21}\hyperref[fig.s21]{(a)-(b)}. 
In order to have a intuitive comprehension of these critical states, we plot the probability distributions of selected eigenstates in Fig.~\ref{fig.s21}\hyperref[fig.s21]{(c)}. 
It reveals that the states remain extended but nonergodic, irrespective of their position relative to the boundaries or centers of energy bands, thus confirming their multifractal nature.

%
\subsection{The effect of large $\lambda$}
\label{secs22}
The localized zone gradually diminishes as $\beta$ increases, which seems to indicate the system enters the localized phase requiring a larger $\lambda$. 
This is confirmed in the left-hand panels of Fig.~\ref{fig.s22}. 
For $\beta=50$ and $\lambda=10$, the upper panel of Fig.~\ref{fig.s22}\hyperref[fig.s22]{(a)-(b)} shows that $D^{(n)}$ for some eigenstates fluctuate away $0$ and $1$, and their probability distributions remain extended but nonergodic, highlighting critical states. Upon increasing $\lambda$ to 100, these states become localized to narrow lattice sites, as depicted in the lower panel of Fig.~\ref{fig.s22}\hyperref[fig.s22]{(a)-(b)}. 
It may be presumed that the system is perpetually localized as long as the potential strength $\lambda$ is large enough. However, this is not the case. 
Take $\beta=10^3$ for example, the upper panel of Fig.~\ref{fig.s22}\hyperref[fig.s22]{(c)} reveals that the fractal dimension $D^{(n)}$ for a large number of eigenstates fluctuates form $0$ and $1$, with only a few exceptions (such as $n=30$), signifying the emergence of many critical states even with $\lambda$ fixed at $100$. 
Most importantly, these critical states persist in an extended but non-ergodic form when $\lambda$ further increases to $1000$, as shown in the blue and green lines in Fig.~\ref{fig.s22}\hyperref[fig.s22]{(d)}.
For $\beta=10^4$, the right-hand panel of Fig.~\ref{fig.s22} shows that the distribution of $D^{(n)}$ for almost all the eigenstates varies from 0 to 1. These critical eigenstates remain extended yet non-ergodic regardless of their proximity to the energy band boundaries or centers, which again corroborates their multifractal nature. 
This behavior stems from the increasing $\beta$, which causes eq. (\ref{2}) to more accurately approximate the potential which alternates between $\pm1$ values following the Fibonacci substitution rule.
The most prominent feature of Fibonacci chain is that all the eigenstates are critical no matter how strong the potential strength $\lambda$ is. 
Consequently, the critical phase remains stable at high values of $\beta$, despite the presence of strong potential strength.
 
\FloatBarrier

\bibliography{main}

\end{document}